\documentclass[nofootinbib,floatfix,twocolumn,superscriptaddress,a4paper,prl]{revtex4}

\usepackage{amsmath}
\usepackage{amssymb}
\usepackage{graphicx}
\usepackage[T1]{fontenc} 
\usepackage{slashed}
\usepackage{xspace}

\newcommand{\vevs}{\textit{vev}s\xspace}

\newcommand{\Br}{\text{Br}}

\newcommand{\UBL}{{\ensuremath{U(1)_{B-L}}}\xspace}
\newcommand{\BL}{{\ensuremath{B-L}}\xspace}

\newcommand{\gBL}[1]{{\ensuremath{g_{BL}^{#1}}}\xspace}
\newcommand{\gmix}{{\ensuremath{\bar{g}}}\xspace}
\newcommand{\gsqsum}{{\ensuremath{g_{\Sigma}^{2}}}\xspace}

\newcommand{\AddrFreiburg}{
Albert-Ludwigs-Universit\"at - Fakult\"at f\"ur Mathematik und Physik,\\
 79104 Freiburg i.\ Br., Germany
 }
\newcommand{\AddrBonn}{%
Bethe Center for Theoretical Physics \& Physikalisches Institut der 
Universit\"at Bonn, \\
 53115 Bonn, Germany
}

\preprint{BONN-TH-2012-27}
\begin{document}

\title{Enhancing $h \to \gamma \gamma$ with staus in SUSY models with extended gauge sector}

\author{L. Basso} \email{lorenzo.basso@physik.uni-freiburg.de}
\affiliation{\AddrFreiburg}

\author{F. Staub}\email{fnstaub@th.physik.uni-bonn.de}
\affiliation{\AddrBonn}


\begin{abstract}
We discuss the impact on the stau masses of the additional D-terms in $U(1)$-extended MSSM models. We show, explicitly for the B-L-SSM,
that these contributions can play a crucial role in the explanation of the enhanced diphoton decay rate of a SM-like Higgs particle around $125$~GeV. Even in the most constrained scenario with universal scalar and gaugino masses, it is possible to obtain a sizable enhancement and, in addition, the correct relic density for the LSP. Furthermore, a lighter $CP$-even scalar that could fit the LEP excess at $98$~GeV is viable.
\end{abstract}
\maketitle

\paragraph*{Introduction.}
Both the ATLAS and CMS collaborations have recently reported the observation of a new resonance compatible with the long sought after Higgs boson~\cite{Observation}. Although the best-fit mass of nearly $125$ GeV as well as the $WW^*$ and $ZZ^*$ rates have been interpreted as consistent with the standard model (SM) and with several of its supersymmetric extensions, the collected data seem to indicate an enhancement of the diphoton rate with respect to the SM yield.
 Certainly, the amount of data currently available does not provide convincing evidences, being the diphoton rate $2\sigma$ away from the SM prediction. Instead of considering this as a mere statistical fluctuation, in this paper we pursue the hypothesis that this is caused by a physical effect. This can be explained in two ways: through a reduction of the $hb\overline{b}$ coupling (leading to a smaller total width enhancing all other partial rates) as in the next-to-minimal supersymmetric standard model (NMSSM)~\cite{Benbrik:2012rm} and/or with extra contributions in the $h\gamma\gamma$ loop. In the latter case, light and largely mixed staus have been immediately recognised to be a viable possibility~\cite{Carena:2011aa,Giudice:2012pf}. This generally requires very large $\tan{\beta}$ values, and it is rather hard to realise in constrained minimal supersymmetric standard model (CMSSM)~\cite{Heng:2012at}. Furthermore, the connection with the explanation of the muon $g-2$ problem has been investigated, both in minimal supersymmetric (SUSY) models~\cite{Carena:2011aa,Giudice:2012pf} (assuming universal soft-slepton masses, light smuons and charginos can achieve the scope), in gauge mediation models~\cite{Sato:2012bf} (where low scale mediators are required), and in their $U(1)$ extensions~\cite{Endo:2011gy} (in  which the described mechanism requires an additional charge for the MSSM doublet Higgses and the $U(1)_{B-L}$ extension is, hence, not considered). 
In the latter models, the dark matter candidate's relic density is also  matched to the observed value.

In this paper we investigate in the $U(1)_{B-L}$ extended MSSM the possibility for a Higgs boson compatible with the recent observations and the mechanisms to enhance its diphoton rate while allowing the $WW^*$ and $ZZ^*$ rates to match the SM predictions. This is achieved with relatively light staus, as in minimal SUSY models, with the only difference being that the new D-terms coming from the $B-L$ sector can further reduce the stau mass entering in the $h\gamma\gamma$ effective interaction (while ensuring a pole mass of $\sim 250$ GeV, compatible with exclusions) \footnote{With pole mass we denote the one-loop corrected mass at $Q=M_{SUSY}=\sqrt{\tilde{t}_1\tilde{t}_2}$, while in the loop, leading to the effective $h\gamma\gamma$ coupling, the running $\overline{\text{DR}}$ tree-level mass at $Q=m_h$ enters.}
 leading this mechanism to work also in the constrained version of the model (as opposed to the CMSSM~\cite{Benbrik:2012rm,Heng:2012at}) and ameliorating the situation in the MSSM where pole stau masses (taken to be equal to the mass entering in the $h\gamma\gamma$ loop) close to the LEP limit of $100$ GeV are required. Such light staus can be looked for at the LHC in the $Z'\to \tilde{\tau}\tilde{\tau}$ channel~\cite{Baumgart:2006pa}. Despite the light stau, we will show that a neutralino DM candidate compatible with observation is viable.
A further viable scenario is represented by the $125$ GeV Higgs boson with an enhanced diphoton rate, accompanied by a lighter state fitting the $2.3\sigma$ excess at $98$ GeV observed at LEP~\cite{LEP-excess}. In this case also a viable DM candidate, either a neutralino or a sneutrino, is possible. Even lighter scalars (down to $10$ GeV~\cite{O'Leary:2011yq}) that would have escaped all searches are possible in this model.
Regarding the muon $g-2$, its explanation requires  low smuon/chargino masses, not possible in our constrained scenario because rather large $m_0$ is needed for spontaneous $B-L$ breaking. If we soften our assumption of strict universality in the scalar sector and allow for non-unified scalar masses, we can choose smaller sfermion masses and as well $g_\mu-2$ can thus be explained.

\paragraph*{The minimal supersymmetric $B-L$ model.}
We describe here the minimal supersymmetric extension of
the SM with a $U(1)_{B-L}$ gauge group~\cite{model}.
The model consists of three generations of matter particles including
right-handed neutrinos which can, for example, be embedded in $SO(10)$
16-plets. We require gauge coupling unification at some grand unified (GUT) scale. Below the GUT scale, the usual MSSM Higgs doublets
are present as well as two fields $\eta$ and $\bar{\eta}$ responsible
for the breaking of the \UBL. Furthermore, $\eta$ is responsible for
generating a Majorana mass term for the right-handed neutrinos and
thus we interpret the \BL charge of this field and of $\bar{\eta}$ as their lepton numbers, and call these fields bileptons since
they carry twice the lepton number of (anti-)neutrinos.  We summarize
the quantum numbers of the chiral superfields with respect to $U(1)_Y
\times SU(2)_L \times SU(3)_C \times \UBL$ in Table~\ref{tab:cSF}.
\begin{table} 
\centering
\begin{tabular}{|c|c|c|c|c|c|} 
\hline \hline 
Superfield & Spin 0 & Spin \(\frac{1}{2}\) & Generations & {\small \(G_{SM} \otimes\, \UBL)\)} \\ 
\hline 
\(\hat{Q}\) & \(\tilde{Q}\) & \(Q\) & 3
 & \((\frac{1}{6},{\bf 2},{\bf 3},\frac{1}{6}) \) \\ 
\(\hat{d}^c\) & \(\tilde{d}^c\) & \(d^c\) & 3
 & \((\frac{1}{3},{\bf 1},{\bf \overline{3}},-\frac{1}{6}) \) \\ 
\(\hat{u}^c\) & \(\tilde{u}^c\) & \(u^c\) & 3
 & \((-\frac{2}{3},{\bf 1},{\bf \overline{3}},-\frac{1}{6}) \) \\ 
\(\hat{L}\) & \(\tilde{L}\) & \(L\) & 3
 & \((-\frac{1}{2},{\bf 2},{\bf 1},-\frac{1}{2}) \) \\ 
\(\hat{e}^c\) & \(\tilde{e}^c\) & \(e^c\) & 3
 & \((1,{\bf 1},{\bf 1},\frac{1}{2}) \) \\ 
\(\hat{\nu}^c\) & \(\tilde{\nu}^c\) & \(\nu^c\) & 3
 & \((0,{\bf 1},{\bf 1},\frac{1}{2}) \) \\ 
\(\hat{H}_d\) & \(H_d\) & \(\tilde{H}_d\) & 1
 & \((-\frac{1}{2},{\bf 2},{\bf 1},0) \) \\ 
\(\hat{H}_u\) & \(H_u\) & \(\tilde{H}_u\) & 1
 & \((\frac{1}{2},{\bf 2},{\bf 1},0) \) \\ 
\(\hat{\eta}\) & \(\eta\) & \(\tilde{\eta}\) & 1
 & \((0,{\bf 1},{\bf 1},-1) \) \\ 
\(\hat{\bar{\eta}}\) & \(\bar{\eta}\) & \(\tilde{\bar{\eta}}\) & 1
 & \((0,{\bf 1},{\bf 1},1) \) \\ 
\hline \hline
\end{tabular} 
\caption{Chiral superfields and their quantum numbers. $G_{SM}$ are the
standard model gauge groups: $G_{SM} = U(1)_Y \otimes\, SU(2)_L \otimes\, SU(3)_c$.}
\label{tab:cSF}
\end{table}

The superpotential is given by
\begin{align} 
\nonumber 
W = & \, Y^{ij}_u\,\hat{u}^c_i\,\hat{Q}_j\,\hat{H}_u\,
- Y_d^{ij} \,\hat{d}^c_i\,\hat{Q}_j\,\hat{H}_d\,
- Y^{ij}_e \,\hat{e}^c_i\,\hat{L}_j\,\hat{H}_d\,\, \\
 & \, \, +\mu\,\hat{H}_u\,\hat{H}_d
+Y^{ij}_{\nu}\,\hat{\nu}^c_i\,\hat{L}_j\,\hat{H}_u\,
- \mu' \,\hat{\eta}\,\hat{\bar{\eta}}\,
+Y^{ij}_x\,\hat{\nu}^c_i\,\hat{\eta}\,\hat{\nu}^c_j\, .
\label{eq:superpot}
\end{align} 
For details about the additional soft SUSY-breaking terms for scalar masses, gaugino masses and scalar interactions, see~\cite{O'Leary:2011yq}.
The extended gauge group breaks to
$SU(3)_C \otimes U(1)_{em}$ as the Higgs and bilepton fields receive
vacuum expectation values (\vevs):
\begin{align} \nonumber
H_d^0 = & \, \frac{1}{\sqrt{2}} \left(\sigma_{d} + v_d  + i \phi_{d} \right),
\,
H_u^0 = \, \frac{1}{\sqrt{2}} \left(\sigma_{u} + v_u  + i \phi_{u} \right),\\ 
\eta
= & \, \frac{1}{\sqrt{2}} \left(\sigma_\eta + v_{\eta} + i \phi_{\eta} \right),
\,
\bar{\eta}
= \, \frac{1}{\sqrt{2}} \left(\sigma_{\bar{\eta}} + v_{\bar{\eta}}
 + i \phi_{\bar{\eta}} \right).
\end{align} 
We define $\tan\beta' = v_{\eta}/v_{\bar{\eta}}$
 in analogy to
the ratio of the MSSM \vevs ($\tan\beta = v_{u}/v_{d}$). \\
The $CP$-odd sector of the MSSM and the $B-L$ decouple at tree-level and the 
masses of the physical pseudoscalars $A^0$ and $A^0_\eta$ are given
 by
 \begin{equation}
 m^2_{A^0} = \frac{2 B_\mu}{\sin2\beta} \thickspace, \hspace{1cm} 
 m^2_{A^0_\eta}  = \frac{2 B_{\mu'}}{\sin2\beta'} \thickspace.
 \end{equation} 
In the scalar sector the gauge kinetic terms induce a
 mixing between the $SU(2)$ doublet Higgs fields and the bileptons. We
 define
  $\gsqsum = \frac{1}{4} (g^2_1+g^2_2+{\gmix}^{2}),
  c_x = \cos(x)$,
  and $s_x = \sin(x)$ ($x=\beta,\beta')$, 
 and use $g_{BL}$ for the $B-L$ gauge coupling and $\gmix$ for the off-diagonal
 gauge coupling induced by kinetic mixing. With this conventions the mass matrix reads at tree level in the 
$(\sigma_d,\sigma_u,\sigma_\eta,\sigma_{\bar{\eta}})$ basis as
\begin{widetext}
\begin{align}
& m^2_{h,T} = \nonumber \\
&\left(\begin{array}{cccc}
m^2_{A^0} s^2_\beta + \gsqsum v^2_u & \,\,
-m^2_{A^0} c_\beta s_\beta - \gsqsum v_d v_u &
 \frac{\gmix \gBL{}}{2}   v_d v_{\eta} &
  -\frac{\gmix \gBL{}}{2} v_d v_{\bar{\eta}} \\
-m^2_{A^0} c_\beta s_\beta  - \gsqsum v_d v_u &
m^2_{A^0} c^2_\beta + \gsqsum v^2_d & \,\,
 - \frac{\gmix \gBL{}}{2}   v_u v_{\eta} &
  \frac{\gmix \gBL{}}{2} v_u v_{\bar{\eta}} \\
\frac{\gmix \gBL{}}{2}   v_d v_{\eta} &
 - \frac{\gmix \gBL{}}{2}   v_u v_{\eta}  &
  m^2_{A^0_\eta} c^2_{\beta'} + \gBL{2} v^2_\eta &
 \,\, -  m^2_{A^0_\eta} c_{\beta'} s_{\beta'} 
  - \gBL{2} v_\eta v_{\bar{\eta}} \\
-\frac{\gmix \gBL{}}{2} v_d v_{\bar{\eta}} &
 \frac{\gmix \gBL{}}{2} v_u v_{\bar{\eta}}  &
\,\, -  m^2_{A^0_\eta} c_{\beta'} s_{\beta'} 
  - \gBL{2} v_\eta v_{\bar{\eta}} &
 m^2_{A^0_\eta} s^2_{\beta'} + \gBL{2} v^2_{\bar{\eta}}
\end{array}
\right).
\end{align}
\end{widetext}
Notice that the two sectors would decouple at tree level if kinetic mixing is
neglected. However, including kinetic mixing results in an enhanced
mass especially of the light MSSM-like Higgs due to the mixing with the bilepton sector~\cite{O'Leary:2011yq}. 
This happens especially near the level crossing when the light bilepton
and the light doublet states are close in mass and can cause a non-negligible doublet 
fraction of the light bilepton. 

\begin{figure}[hbt]
\includegraphics[width=0.9\linewidth]{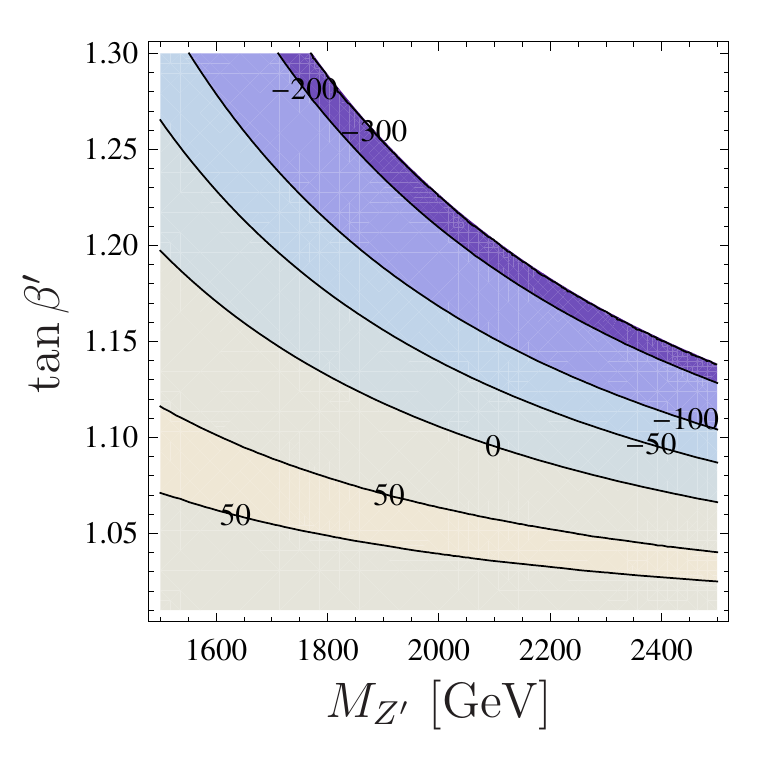} \\
\includegraphics[width=0.9\linewidth]{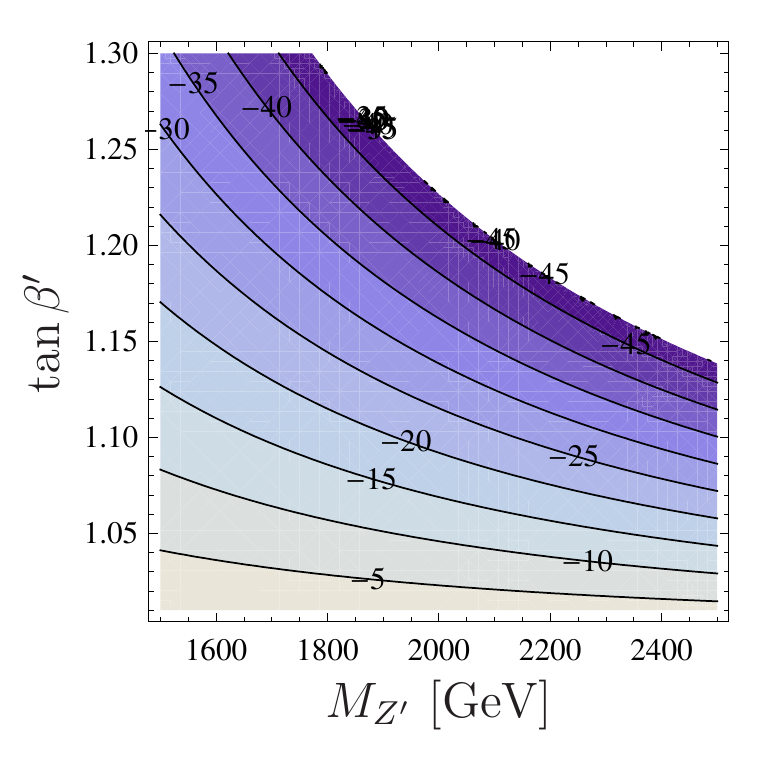}
\caption{Impact of the D-term contributions due to the extended gauge sector on the light stau (upper plot) 
and stop mass (lower plot). The contour lines show the difference to the expectation in the MSSM. 
The input parameters are $\tan\beta =40$, $2 m^{33,2}_l = m^{33,2}_e = 10^5~\text{GeV}^2$,
$2 m_q^{33,2} = m_u^{33,2} = 4\cdot 10^5~\text{GeV}^2$, $\mu = 500$, $T_u^{33} = T_e^{33}=1$~TeV. For these
points the masses in the MSSM would be $m_{\tilde{\tau}_1} = 317$~GeV and $m_{\tilde{t}_1} = 575$~GeV.
}
\label{fig:DeltaM}
\end{figure}

The other matrix important for our consideration here is the one for charged
sleptons. This mass matrix reads, in the 
 \( \left(\tilde{e}_{L}, \tilde{e}_{R}\right) \) basis, as
\begin{equation} 
m^2_{\tilde{e}} = \left( 
\begin{array}{cc}
m_{LL} & \frac{1}{\sqrt{2}} \Big(v_d T_{e}  - v_u \mu^* Y_{e} \Big)\\ 
\frac{1}{\sqrt{2}} \Big(v_d T^\dagger_{e}  - v_u \mu Y^\dagger_{e} \Big)
 & m_{RR}\end{array} 
\right) ,
\end{equation} 
with
\begin{align} 
m_{LL} & =  m_{L}^{2} +\frac{v_d^2}{2} Y^\dagger_{e} Y_{e} + \frac{1}{8} \Big(
 (g_{1}^{2}  -g_{2}^{2})(v_{d}^{2}- v_{u}^{2})
 \nonumber \\ &  +2 g_{BL}^{2} (v_{\eta}^{2}  - v_{\bar{\eta}}^{2})\Big)
 {\bf 1},
\label{eq:mLL11}
\end{align} 
\begin{align} 
m_{RR} & = m_{E}^{2}+\frac{v_d^2}{2} Y_{e} Y^\dagger_{e}
 + \frac{1}{8} \Big(2g_{1}^{2}
 (v_{u}^{2}- v_{d}^{2}) \nonumber \\ &  -2\Big(2 g_{BL}^{2} \Big)(v_{\eta}^{2}
 - v_{\bar{\eta}}^{2}) \Big){\bf 1}.
\end{align} 
In these equations we have suppressed  the terms coming from kinetic mixing, since the latter
 plays a smaller role here than for the Higgs particles. Of course, in our 
numerical studies all terms are taken into account. 
In general, we can parametrize the D-term contributions as a function of the $Z'$ mass, $M_{Z'}$, and of $\tan\beta'$ as
\begin{equation}
\frac{Q^{B-L}}{2} \frac{M_{Z'} (\tan^2\beta' -1)}{1 + \tan^2 \beta'}.
\end{equation}
Obviously, the D-term contributions from the \BL sector are larger for the sleptons than for the squarks by a factor of $3$ due to the different $B-L$ charges. Because of this, the effect discussed in the following is much more pronounced in the slepton sector and it is possible to get large effects for staus while keeping the impact on the squarks under control. 
This is depicted in Fig.~\ref{fig:DeltaM}, where we show the mass difference of staus (top-panel) and stops (bottom-panel) in the B-L-SSM in comparison to the MSSM expectations for a fixed set of SUSY-scale parameters. 

\begin{figure}[hbt]
\includegraphics[width=0.9\linewidth]{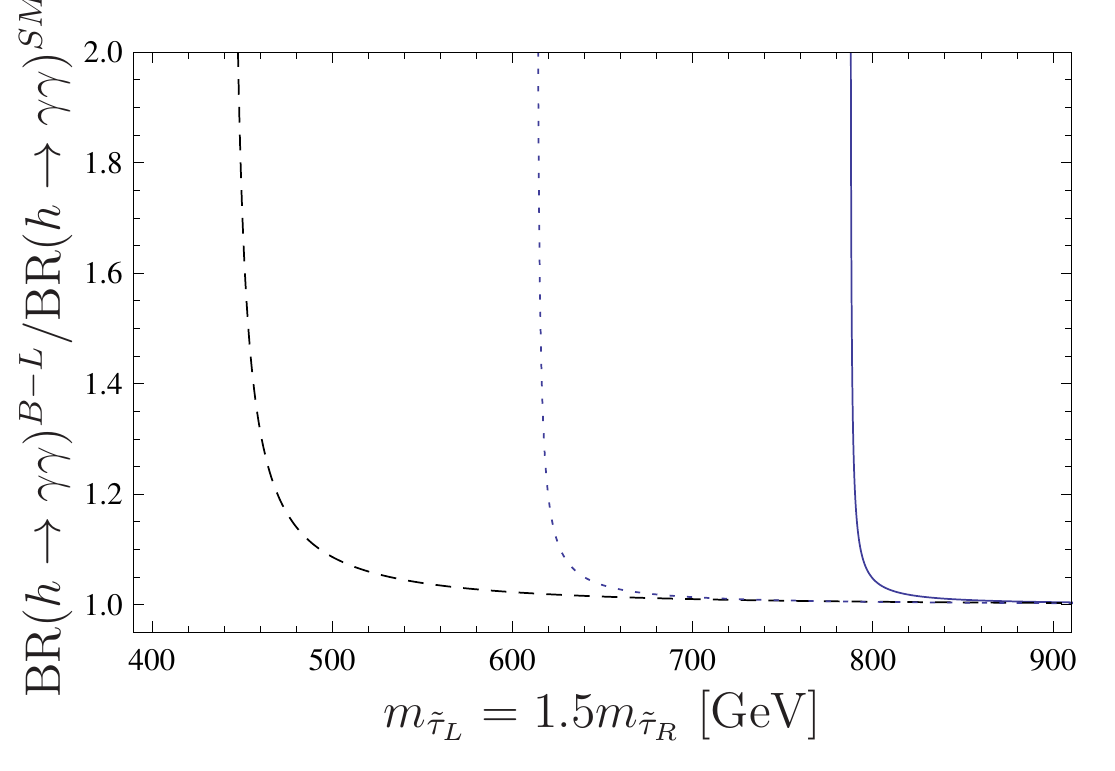} 
\caption{$\Br(h\to \gamma \gamma)^{B-L}/\Br(h\to \gamma\gamma)^{SM}$ as a function of $m_{\tilde{\tau}_L} = 1.5 m_{\tilde{\tau}_R} $. 
The dashed line is the SM limit without additional D-term contributions, the solid line shows the 
case for $\tan\beta' = 1.5$ and the small-dashed one for $\tan\beta' = 1.25$. The other parameters have been set to $A_\tau = 1$~TeV,
$\mu = 1.5$~TeV,  and $M_{Z'} = 2.7$~TeV.  }
\label{fig:MSSMvsBL}
\end{figure}

\paragraph*{$h \to \gamma \gamma$ in the B-L-SSM.}

To start our discussion let us briefly review the partial decay width of the Higgs boson $h$ into two photons within the MSSM and its singlet extensions.
This can be written as (see, e.g.,~\cite{Djouadi:2005gi})
\begin{align}
\label{eq:decaywidth}
&\Gamma_{h \rightarrow\gamma\gamma} 
 = \frac{G_\mu\alpha^2 m_{h}^3}
{128\sqrt{2}\pi^3} \bigg| \sum_f N_c Q_f^2 g_{h ff} A_{1/2}^{h} 
(\tau_f) + g_{h WW} A_1^{h} (\tau_W) \nonumber \\ 
& \hspace*{0.2cm} + \frac{m_W^2 g_{h H^+ H^-} }{2c_W^2 
m_{H^\pm}^2} A_0^h(\tau_{H^\pm})  +  \sum_{\chi_i^\pm} \frac{2 m_W}{ m_{\chi_i^\pm}} g_{h \chi_i^+ 
\chi_i^-} A_{1/2}^{h} (\tau_{\chi_i^\pm}) \nonumber \\ 
& \hspace*{0.2cm}
+\sum_{\tilde e_i} \frac{ g_{h \tilde e_i \tilde e_i} }{ 
m_{\tilde{e}_i}^2} \,  A_0^{h} (\tau_{ {\tilde e}_i}) +\sum_{\tilde q_i} \frac{ g_{h \tilde q_i \tilde q_i} }{ 
m_{\tilde{q}_i}^2} \, 3 Q_{\tilde q_i}^2 A_0^{h} (\tau_{ {\tilde q}_i})  \bigg|^2\,,
\end{align}
corresponding to the contributions from charged SM fermions, $W$ bosons, charged Higgs, charginos, charged sleptons and squarks, respectively.
The amplitudes $A_i$ at lowest order for the 
spin--1, spin--$\frac{1}{2}$  and spin--0 particle contributions,
can be found for instance in Ref.~\cite{Djouadi:2005gi}.
 $g_{hXX}$ denotes the coupling between the Higgs boson and the particle in the loop and $Q_X$ is its electric charge.
In the SM, the largest contribution is given by the $W$-loop, while the top-loop leads to a small reduction of the decay rate. In the MSSM, it is possible to get large contributions due to sleptons and squarks, although it is difficult to realize such a scenario in a constrained model with universal sfermion masses \cite{Carena:2011aa,Benbrik:2012rm}. 
In singlet or triplet extension of the MSSM also the chargino and charged Higgs can enhance the loop significantly~\cite{SchmidtHoberg:2012yy}. 
However, this is only possible for large singlet couplings which lead to a cut-off well below the GUT scale. In contrast, it is possible to enhance the diphoton ratio in the B-L-SSM due to light staus even in the case of universal boundary conditions at the GUT scale. We show this by calculating explicitly the contributions of the stau:
\begin{align}
 & A(\tilde{\tau}) = \frac{1}{3} \frac{\partial \text{det} m_{\tilde{\tau}}^2}{\partial \log v}  \\ 
  \simeq& -\frac{2}{3} \frac{2 m_{\tau}^2 (A_\tau - \mu \tan\beta)^2}{(m_E^2 + D_R)(m_L^2 + D_L) + m_{\tau}^2 \mu \tan\beta(2 A_\tau - \mu \tan\beta)}\, .
\end{align}
Here, $D_L$ and $D_R$ represent the D-term contributions of the left- and right-handed stau and we have neglected sub-leading contributions. Given that
 $2 A_\tau < \mu \tan\beta$, for fixed values of the other parameters, $D_R$ and $D_L$ can be used to enhance the $\gamma\gamma$ rate by suppressing the denominator. This is shown in Fig.~\ref{fig:MSSMvsBL}: depending on $\tan\beta'$, it is possible to obtain enhanced branching ratios (BR) of the SM-like Higgs into photons even for much heavier soft masses than possible in the MSSM.

We turn now to a fully numerical analysis to demonstrate the mechanism to reduce the stau mass in comparison to the stop mass (to enhance the Higgs to diphoton rate) discussed above.
For this purpose, we used the {\tt SPheno} version \cite{spheno} created with {\tt SARAH} \cite{sarah} for the \mbox{B-L-SSM}. This spectrum calculator performs a two-loop RGE evaluation and calculates the mass spectrum at one loop. In addition, it calculates the decay widths and BRs of all SUSY and Higgs particles as well as observables like $\Delta a_\mu$. We will discuss the most constrained scenario with a universal scalar mass $m_0$, a universal gaugino mass $M_{1/2}$ and trilinear softbreaking couplings proportional to the superpotential coupling ($T_i = A_0 Y_i$) at the GUT scale. Other input parameters are $\tan\beta$, $\tan\beta'$, $M_{Z'}$ and $Y_x$. $Y_\nu$ is fixed by neutrino data and is, thus, tiny and negligible for our purposes. For a more detailed discussion of the code we refer to~\cite{O'Leary:2011yq}. \\
\begin{figure}[hbt]
\begin{tabular}{r}
\includegraphics[width=0.765\linewidth]{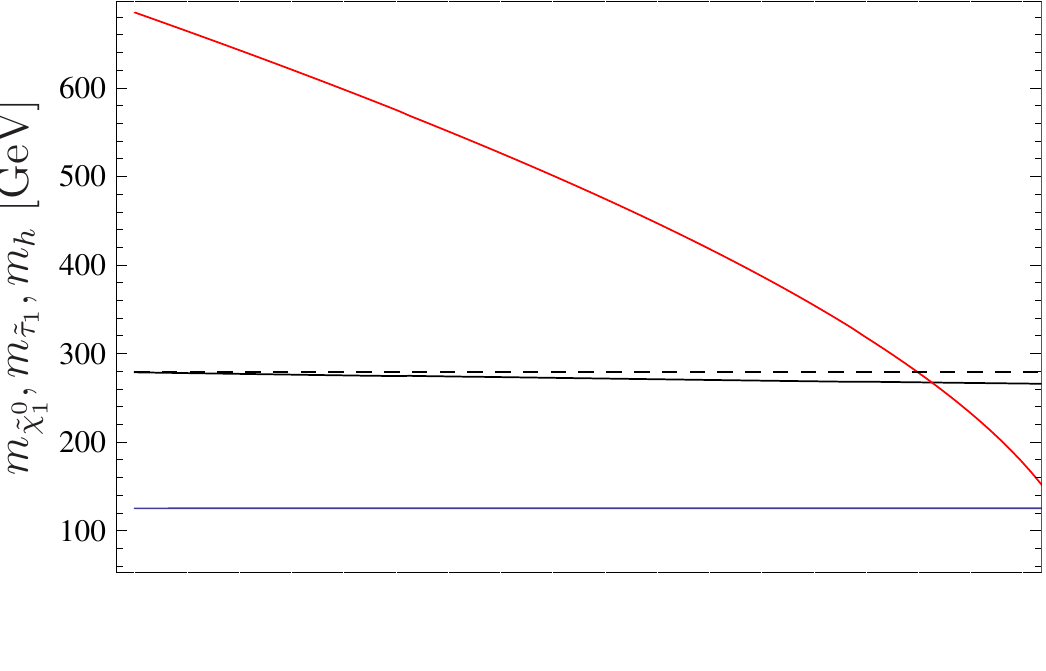} \vspace{-0.75cm}\\
\includegraphics[width=0.76\linewidth]{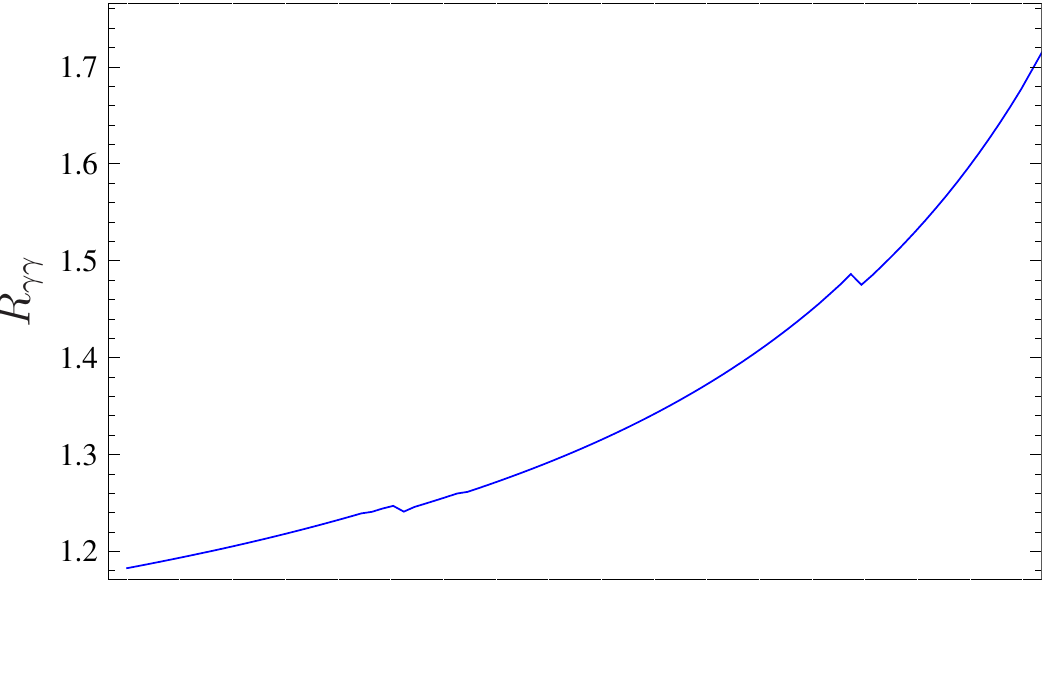} \vspace{-0.81cm}\\
\includegraphics[width=0.757\linewidth]{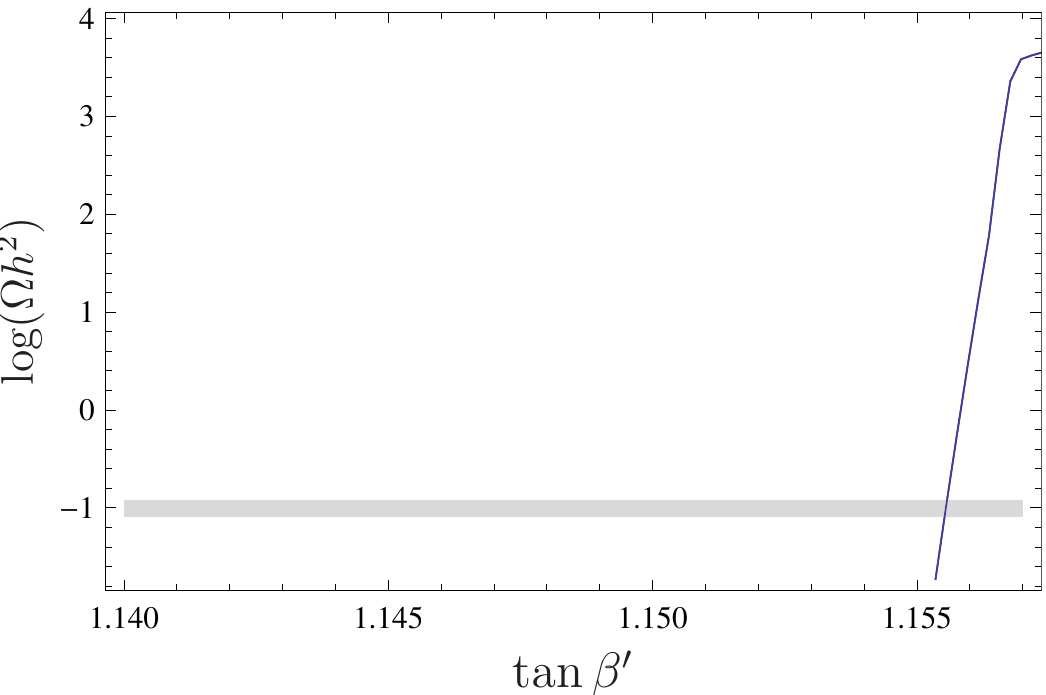} \vspace{-0.00cm}
\end{tabular}
\caption{(Top plot) The mass of the SM-like Higgs [bottom(blue line)], of the stau [middle(black) line, where the dashed line represents a reference unchanged value] and of the sneutrino [top(red) line]; (middle plot) the diphoton branching ratio; (bottom plot) the neutralino relic density as a function of $\tan\beta'$. The other parameters have been chosen as $m_0=673$~GeV, $M_{1/2} = 2220$~GeV, $\tan\beta=42.2$, $A_0 = -1842.6$, $M_{Z'} = 2550$~GeV, $Y_x = {\bf 1} \cdot 0.42$}
\label{fig:varTBp}
\end{figure}
In Table~\ref{tab:benchmark} we have collected two possible scenarios that provide a SM-like Higgs particle in the mass range preferred by LHC results with an enhanced diphoton rate. In the first point, the lightest $CP$-even scalar eigenstate is the SM-like Higgs boson while the light bilepton is roughly twice as heavy. In Fig.~\ref{fig:varTBp} we show that all the features arise from the extended gauge sector: it is sufficient to change only $\tan\beta'$ to obtain an enhanced diphoton signal $R^1_{\gamma \gamma}\equiv \frac{\left[ \sigma (gg\to h_1) \cdot BR(h_1\to \gamma\gamma)\right]_{B-L}}{\left[ \sigma (gg\to h_1) \cdot BR(h_1\to \gamma\gamma)\right]_{SM}}$ and the correct dark matter relic density while keeping the mass of the SM-like Higgs nearly unchanged. The dark matter candidate in this scenario is the lightest neutralino, that is mostly a bileptino (the superpartner of the bileptons). The correct abundance for $\tan\beta' \simeq 1.156$ is obtained due to a co-annihilation with the light stau. 
In the second point, the SM-like Higgs is accompanied by a light scalar around $98$~GeV which couples weakly to the SM gauge bosons, compatibly with the LEP excess~\cite{LEP-excess}. In this case, the LSP is a $CP$-odd sneutrino which annihilates very efficiently due to the large $Y_x$. This usually results in a small relic density. To get an abundance which is large enough to explain the dark matter relic, the mass of the sneutrino has to be tuned below $m_W$~\cite{Basso:2012gz}. This can be achieved by slightly increasing $\tan{\beta '}$ and by tuning the Majorana Yukawa couplings $Y_x$, that tends to increase the SM-like Higgs mass for the given point. It is worth mentioning that a neutralino LSP with the correct relic density in the stau co-annihilation region can also be found in this scenario. 
Notice that both points yield rates consistent with observations in the $WW^*/ZZ^*$ channels (measured at the LHC) and in the $b\overline{b}$ channel, as recently claimed by Tevatron~\cite{Aaltonen:2012qt} (being $c_{hZZ}\sim 1$).

\begin{table}[h]
\begin{tabular}{|c|cc|}
\hline \hline
& \parbox{1.5cm}{\centering Point I} 
& \parbox{1.5cm}{\centering Point II} \\
\hline
\hline
$m_{h_1}$~[GeV] & 125.2   & 98.2    \\
$m_{h_2}$~[GeV] & 186.9   & 123.0    \\
$m_{\tilde{\tau}}$ [GeV] & 267.0 & 237.3 \\
\hline
doublet fr. [\%] & 99.5 & 8.7  \\
bilepton fr. [\%]& 0.5 &  91.3 \\
\hline
$c_{h_1 g g}$ &  0.992 & 0.087   \\
$c_{h_1 Z Z}$ &  1.001 & 0.085   \\
$c_{h_2 g g}$ &  0.005 & 0.911   \\
$c_{h_2 Z Z}$ &  0.005 & 0.921   \\
\hline
$\Gamma(h_1) $~[MeV]                            & 4.13 & 0.22   \\
$R^1_{ \gamma \gamma}$ & 1.57 &  0.085   \\
$R^1_{b \overline{b}}$ & 1.03 & 0.089  \\
$R^1_{WW^*}$ & 0.98 & 0.05  \\
\hline
$\Gamma(h_2) $~[MeV]                            & 4.8 & 3.58  \\
$R^2_{ \gamma \gamma}$ & 0.005 &  1.79   \\
$R^2_{b \overline{b}}$ & 0.006 & 0.95  \\
$R^2_{WW^*}$ & 0.01 & 0.88  \\
\hline
LSP mass  ~[GeV]           &  $253.9$    &  $82.9$   \\
$\Omega h^2 $              &  $0.10$    &  $10^{-2}$   \\
\hline
\hline                    
\end{tabular}
\caption{ The input parameter used: Point I: $m_0 = 673$~GeV , $M_{1/2} = 2220$~GeV, $A_0 = -1842$~GeV, $\tan\beta=42.2$, $\tan\beta'=1.1556$, $M_{Z'} = 2550$~GeV, $Y_x = {\bf 1} \cdot 0.42$ (neutralino LSP). Point II: $m_0 = 742$~GeV , $M_{1/2} = 1572$~GeV, $A_0 = 3277$~GeV, $\tan\beta=37.8$, $\tan\beta'=1.140$, $M_{Z'} = 2365$~GeV, $Y_x=\text{diag}(0.40,0.40,0.13)$ (CP-odd sneutrino LSP). $c_{SVV}$ denotes the coupling squared of the Higgs fields to vector bosons normalized to the SM values. } 
\label{tab:benchmark}
\end{table}

\paragraph*{Conclusions.}
In summary, in this letter we have discussed the impact of an extended gauge sector on the diphoton rate of a SM-like Higgs boson compatible with recent observations. It has been shown that it is possible to enhance BR($h \to \gamma \gamma$) with light staus even
in a constrained GUT version of the B-L-SSM model, due to its extra D-terms.
However, the impact of these terms on the squarks which unify with the other scalars at the GUT scale is marginal, so that they remain heavy enough to explain the Higgs mass of $125$~GeV. In addition, the bounds on the dark matter relic density can be satisfied by either a neutralino or a sneutrino LSP. Finally,  a lighter scalar with a mass of $98$~GeV may also be present beside the SM-like Higgs. Due to the mixing with the doublet state, this scalar couples weakly to the SM particles and could have caused the excess measured at LEP.

We thank Shaaban Khalil, Stefano Moretti, Shoaib Munir, Ben O'Leary, Werner Porod, and Giovanni Marco Pruna for stimulating and fruitful discussions and for reading the manuscript. LB is supported by the
Deutsche Forschungsgemeinschaft through the Research Training Group grant
GRK\,1102 \textit{Physics of Hadron Accelerators}.

\end{document}